\documentclass[12pt]{iopart}
\usepackage{iopams}

\usepackage[utf8]{inputenc}
\usepackage{graphicx}
\usepackage[colorlinks=true,citecolor=blue]{hyperref}
\usepackage{units}
\usepackage{verbatim} 

\newcommand{\kBT}{k_\text{B}T}
\newcommand{\text}[1]{\mathsf{#1}}

\def\kB {k_\text{B}}

\usepackage{blindtext}

\begin{document}
\title{High-efficiency thermal switch based on topological Josephson junctions}
\author{Bj\"orn Sothmann$^{1,2}$, Francesco Giazotto$^3$, Ewelina M. Hankiewicz$^2$}
\address{$^1$Theoretische Physik, Universität Duisburg-Essen and CENIDE, D-47048 Duisburg, Germany}
\address{$^2$Institute for Theoretical Physics and Astrophysics, University of Würzburg, Am Hubland, 97074 Würzburg, Germany}
\address{$^3$NEST, Istituto Nanoscienze-CNR and Scuola Normale Superiore, I-56127 Pisa, Italy}
\eads{\mailto{bjoerns@thp.uni-due.de},\\
\mailto{francesco.giazotto@sns.it},\\
\mailto{ewelina.hankiewicz@physik.uni-wuerzburg.de}}
\date{\today}

\begin{abstract}
We propose theoretically a thermal switch operating by the magnetic-flux controlled diffraction of phase-coherent heat currents in a thermally biased Josephson junction based on a two-dimensional topological insulator. For short junctions, the system shows a sharp switching behavior while for long junctions the switching is smooth. Physically, the switching arises from the Doppler shift of the superconducting condensate due to screening currents induced by a magnetic flux. We suggest a possible experimental realization that exhibits a relative temperature change of 40\% between the on and off state for realistic parameters. This is a factor of two larger than in recently realized thermal modulators based on conventional superconducting tunnel junctions.
\end{abstract}

\submitto{\NJP}
\maketitle

\section{Introduction}
Transport of electrical charge is at the heart of modern computers which are based on building blocks like diodes, transistors and switches. Over the past decades, assembling an ever increasing number of these constituents on computer chips has led to an exponential growth of computational power. At the same time, 
the biggest problem  to keep Moore's law, the doubling of the number of transistors on a chip every two years, working is connected with the heat production due to Joule heating.  As a consequence, proper thermal management and active cooling are required.

This raises the question whether it is possible to build alternative logical devices based on thermal rather than electrical transport, thus overcoming the problems associated with Joule heating. A promising route towards this goal is phase-coherent caloritronics in nanoscale superconducting circuits~\cite{martinez-perez_coherent_2014}. Apart from potential applications in thermal logic~\cite{li_colloquium:_2012}, phase-dependent heat currents can also have an impact on the cooling of electronic circuits~\cite{giazotto_opportunities_2006}, radiation detection~\cite{giazotto_opportunities_2006} as well as  quantum information processing~\cite{ladd_quantum_2010,spilla_measurement_2014}. The key ingredient of phase-coherent superconducting caloritronics is the fact that heat currents between superconductors depend on the phase difference between superconducting order parameters. This effect has been predicted theoretically for tunnel junctions ~\cite{maki_entropy_1965,maki_entropy_1966} and superconducting point contacts of arbitrary transmission~\cite{zhao_phase_2003,zhao_heat_2004}. It arises from the interference between quasiparticles carrying heat but no phase information and Cooper pairs carrying phase information but no heat~\cite{guttman_phase-dependent_1997,guttman_interference_1998}.

Only very recently phase-dependent heat currents have been observed experimentally in a superconducting quantum interference device (SQUID)~\cite{giazotto_josephson_2012}.
Subsequently, the diffraction of heat currents by a magnetic flux through an extended Josephson junction has been demonstrated in full analogy to the Fraunhofer pattern of electrical Josephson currents~\cite{giazotto_coherent_2013,martinez-perez_quantum_2014}.
Furthermore, thermal rectifiers~\cite{martinez-perez_efficient_2013,giazotto_thermal_2013,fornieri_electronic_2015} based on normal metal-insulator-superconductor (NIS) junctions have been realized experimentally~\cite{martinez-perez_rectification_2015} as well as a fully balanced heat modulator based on a double-loop SQUID~\cite{fornieri_nanoscale_2016}.
In addition, various proposals to realize refrigerators~\cite{camarasa-gomez_superconducting_2014,solinas_microwave_2016}, heat engines~\cite{hofer_quantum_2016,marchegiani_self-oscillating_2016}, thermometers~\cite{giazotto_ferromagnetic-insulator-based_2015}, heat valves~\cite{strambini_proximity_2014}, and thermal transistors~\cite{giazotto_proposal_2014,fornieri_negative_2016}, have been made. However, phase-dependent thermal transport in topological insulators and topological superconductors has been an unexplored territory so far.

Two-dimensional topological insulators are a novel exotic state of matter that has recently received a lot of interest. They are characterized by an insulating bulk such that conduction can occur only along two helical edge channels~\cite{bernevig_quantum_2006,konig_quantum_2007}. When coupled to a conventional BCS superconductor, the possibility to generate spin-triplet $p$-wave pairing in the topological insulator arises due to strong spin-orbit coupling.
This, in turn, can lead to the formation of topologically protected, gapless Andreev bound states in a superconductor-topological insulator Josephson junction~\cite{fu_superconducting_2008,tkachov_helical_2013}. Topological Andreev bound states manifest themselves via a fractional Josephson effect which is $4\pi$-periodic in the phase difference across the junction rather than being $2\pi$-periodic as in conventional Josephson junctions~\cite{fu_josephson_2009,olund_current-phase_2012}. 
In addition, topological Andreev bound  states also give rise to a characteristic phase-dependence of the thermal conductance of a temperature-biased Josephson junction~\cite{sothmann_fingerprint_2016}.
Gapless Andreev bound states are closely related to Majorana fermions which can be important building blocks for future topological quantum computers allowing for the realization of qubits topologically protected against decoherence~\cite{aasen_milestones_2016}.

Experimentally, topological Josephson junctions have been realized based on HgTe/CdTe quantum wells. 
The oscillations of the supercurrent with an applied magnetic flux resembling that of a SQUID rather than a conventional Fraunhofer pattern have provided evidence for superconducting transport along the edge channels~\cite{hart_induced_2014}.
More recently, signatures of $4\pi$-periodic Josephson currents have been observed in the AC Josephson effect~\cite{deacon_josephson_2016,bocquillon_gapless_2017,wiedenmann_4-periodic_2016}.
Furthermore, the observation of finite momentum pairing has been reported~\cite{hart_controlled_2017}.
In addition, Josephson junctions based on the two-dimensional topological insulator InAs/GaSb have been realized experimentally. Here, a transition of the quantized conductance from $2e^2/h$ to $4e^2/h$ due to Andreev reflection has been measured~\cite{knez_andreev_2012} as well as evidence of supercurrents carried by edge channels~\cite{pribiag_edge-mode_2015}.

\begin{figure}
	\centering\includegraphics[width=.75\columnwidth]{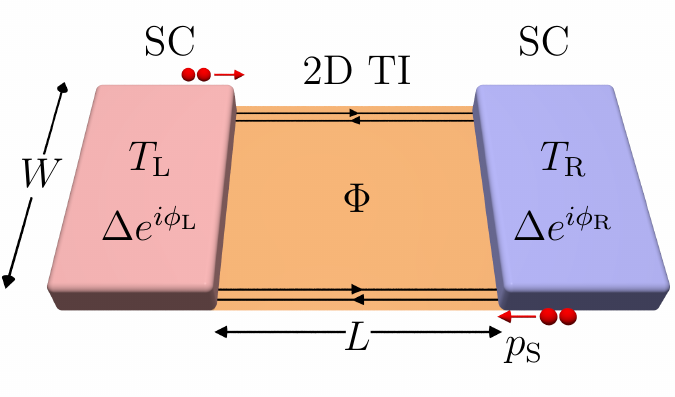}
	\caption{\label{fig:setup_2D}Sketch of a Josephson junction of length $L$ and width $W$ based on a two-dimensional topological insulator. Transport occurs along helical edge states on the two sides of the topological insulator. A magnetic field perpendicular to the plane of the junction gives rise to a magnetic flux $\Phi$ through the junction. Furthermore, it yields a finite Cooper pair momentum $p_S$ due to screening currents inside the superconducting contacts.}
\end{figure}

In this paper, we propose a temperature-biased Josephson junction based on the edge states of a two-dimensional topological insulator as a thermal switch that can be controlled by a weak magnetic field, see Fig.~\ref{fig:setup_2D}. We demonstrate that the switch is very efficient and can achieve a large temperature difference between the on and off state. The effect is robust against the presence of additional transport via the lowest bulk modes.
The switching behavior is controlled by the junction length. While short junctions exhibit a sharp switching behavior, the switching is more smooth for long junctions. 
The underlying physical mechanism is the Doppler shift of the superconducting condensate due to screening currents flowing in response to the magnetic flux~\cite{tkachov_magnetic_2004,tkachov_quantum_2015}. At a critical flux value, the Doppler shift closes the superconducting gap induced in the topological insulator, thereby lifting the exponential suppression of thermal conductance in the superconducting state.
We emphasize that in order to realize an abrupt switching behavior, transport via edge channels is crucial. It guarantees that the gap closes for all edge modes at exactly the same flux, in stark contrast to transport via bulk modes where each conduction channel closes at a different flux.

The paper is organized as follows. In Sec.~\ref{sec:model}, we present our theoretical model of a topological Josephson junction based on the Bogoliubov-de Gennes formalism. We discuss our results in Sec.~\ref{sec:results}, analyzing the transmission function, the thermal conductance as well as a possible experimental realization of a thermal switch. Finally, we draw conclusions in Sec.~\ref{sec:conclusions}.

\section{\label{sec:model}Model}
We consider a Josephson junction of length $L$ and width $W$ based on helical edge channels of a two-dimensional topological insulator and subject to a homogeneous perpendicular magnetic field $B$, cf. Fig.~\ref{fig:setup_2D}. The two superconducting electrodes $i=\text{L,R}$ are kept at different temperatures $T_i$ such that a temperature bias $\Delta T=T_\text{L}-T_\text{R}$ across the junction gives rise to a heat current which in linear response can be written as $J^h=\kappa \Delta T$ with the thermal conductance $\kappa$.

The two edges of the topological insulator are described by the Bogoliubov-de Gennes Hamiltonian
\begin{equation}
	H_\text{BdG}=\left(\begin{array}{cc} h(x) & \rmi\sigma_y \Delta(x) \\ -\rmi\sigma_y \Delta^*(x) & -h^*(x) \end{array}\right),
\end{equation}
where $\Delta(x)$ denotes the superconducting pair potential induced in the topological insulator via the proximity effect of the superconducting electrodes. We assume that at the junction, the pair potential changes on a length scale much shorter than the superconducting coherence length $\xi_0=\frac{\hbar v_\text{F}}{\Delta}$ where $v_\text{F}$ is the Fermi velocity. This allows us to approximate the pair potential by a step function, $\Delta(x)=\Delta \rme^{\rmi\phi_\text{L}}\Theta(-x-L/2)+\Delta \rme^{\rmi\phi_\text{R}}\Theta(x-L/2)$, and neglect proximity effects within the junction that would require a self-consistent determination of the pair potential.

The single-particle Hamiltonian describing the upper and lower helical edge channels is given by
\begin{equation}
	h(x)=v_\text{F}\sigma_x\left(-\rmi\hbar \partial_x \pm \frac{p_S}{2}\right)-\mu.
\end{equation}
Here, $\sigma_x$ is the Pauli matrix acting in spin space while $\mu$ is the chemical potential. $p_S=\frac{\pi\xi_0\Delta}{v_\text{F}L} \frac{\Phi}{\Phi_0}$ denotes the absolute value of the Cooper pair momentum that arises from screening currents that flow in response to an externally applied magnetic field, cf. Ref~\cite{tkachov_quantum_2015}. 
The magnetic flux through the junction is given by $\Phi=BLW$ while $\Phi_0$ is the magnetic flux quantum.
Due to the presence of a magnetic field the phase difference $\phi=\phi_\text{R}-\phi_\text{L}$ across the junction  depends on coordinates. We assume that the junction is smaller than the Josephson penetration depth (this is in general a good approximation for nanoscale junctions). In this case, screening effects due to the magnetic field generated by the Josephson current can be neglected and we have
\begin{equation}
	\phi=\frac{2\pi y}{W}\frac{\Phi}{\Phi_0}+\phi_0,
\end{equation}
where $\phi_0$ is the phase difference for $y=0$. We remark that the width of the junction determines the connection between magnetic field $B$ and magnetic flux $\Phi$ but does not affect the transport properties otherwise as long as the junction is wide enough to have a negligible overlap between edge states on opposite sides.

\begin{figure}
	\centering\includegraphics[width=.75\columnwidth]{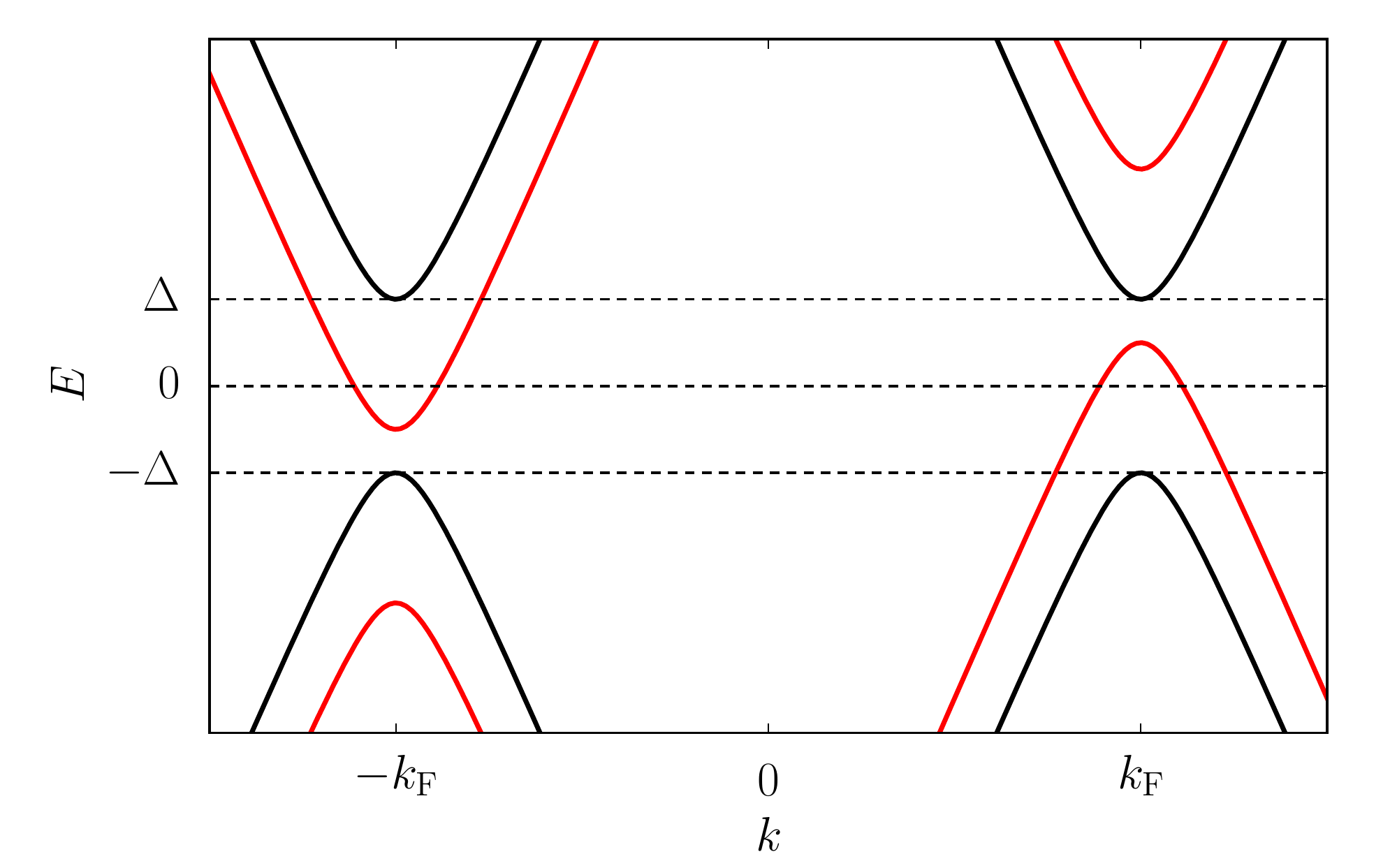}
	\caption{\label{fig:dispersion}Dispersion relation of a two-dimensional topological insulator proximity-coupled to a conventional BCS superconductor. An externally applied magnetic field gives rise to a Doppler shift of the left- and right-moving branches. If the shift is sufficiently large, the proximity-induced gap of the system can be closed, cf. the red lines.}
\end{figure}
The corresponding spectrum inside the superconducting part of the junction is shown in Fig.~\ref{fig:dispersion}. Importantly, due to the Doppler shift, the energies of left and right movers are shifted relative to each other by $v_\text{F}p_S$. If the Doppler shift becomes larger than $2\Delta/v_\text{F}$, the gap of the system closes. We remark that since we assume a proximity-induced superconductivity, there can still be a finite superconducting pairing since the bulk superconductor with a much larger gap $\Delta_0$ provides the necessary Cooper pairs.

The eigenfunctions describing left-moving electron-like quasiparticles, right-moving hole-like quasiparticles, right-moving electron-like quasiparticles and left-moving hole-like quasiparticles in the superconducting region $i=\text{L,R}$ are given by
\begin{eqnarray}
	\psi_1&=(u_-,u_-,-\rme^{-\rmi\phi_i}v_-,\rme^{-\rmi\phi_i}v_-)^T\rme^{+\rmi k_e x},\\
	\psi_2&=(v_-,v_-,-\rme^{-\rmi\phi_i}u_-,\rme^{-\rmi\phi_i}u_-)^T\rme^{+\rmi k_h x},\\
	\psi_3&=(u_+,-u_+,\rme^{-\rmi\phi_i}v_+,\rme^{-\rmi\phi_i}v_+)^T\rme^{-\rmi k_e x},\\
	\psi_4&=(v_+,-v_+,\rme^{-\rmi\phi_i}u_+,\rme^{-\rmi\phi_i}u_+)^T\rme^{-\rmi k_h x},
\end{eqnarray}
respectively, where
\begin{eqnarray}
	u_\pm=\frac{1}{2}\left(1+\frac{\sqrt{\omega_\pm^2-|\Delta|^2}}{\omega_\pm}\right),\\
	v_\pm=\frac{1}{2}\left(1-\frac{\sqrt{\omega_\pm^2-|\Delta|^2}}{\omega_\pm}\right),
\end{eqnarray}
with $\omega_\pm=\omega\pm \frac{v_\text{F}p_S}{2}$ and $k_{e,h}$ denotes the wave vector of electron- and hole-like quasiparticles. Similarly, in the central part of the junction, the eigenfunctions are given by the above expressions with $\Delta=0$.

Let us now consider the case of an electron-like quasiparticle above the superconducting gap incoming from the left. It gives rise to left- and right-moving electrons and holes in the central region of the junction as well as to transmitted and reflected electron- and hole-like quasiparticles. By demanding the continuity of the wave function at the interfaces, we can determine the transmission amplitudes from which the transmission probabilities of quasiparticles follow immediately. We find that the transmission probability for a quasiparticle from the left to the right along the upper edge is given by
\begin{eqnarray}
	\mathcal T_\text{u,R}=&\frac{\omega_-^2-\Delta^2}{\omega_-^2-\Delta^2\cos^2\left(\frac{\phi_u}{2}-\frac{\omega_-L}{\Delta\xi_0}\right)}\Theta(|\omega_-|-\Delta)\\
	&+\frac{\omega_+^2-\Delta^2}{\omega_+^2-\Delta^2\cos^2\left(\frac{\phi_u}{2}+\frac{\omega_+L}{\Delta\xi_0}\right)}\Theta(|\omega_+|-\Delta).
\end{eqnarray}
We omit here Andreev bound state transmission functions, since only quasiparticles above the superconducting gap give rise to the thermal transport. 
The general form of the transmission function is similar to that obtained for a Josephson junction based on surface states of a three-dimensional topological insulator~\cite{sothmann_fingerprint_2016}. We remark that the energy-dependent term inside the cosine arises from the energy dependence of the electron and hole wave vectors and reflects an additional phase picked up by an electron-hole pair making a roundtrip through the junction.
Similarly, for a right-moving particle along the lower edge, we find
\begin{eqnarray}
	\mathcal T_\text{l,R}=&\frac{\omega_+^2-\Delta^2}{\omega_+^2-\Delta^2\cos^2\left(\frac{\phi_l}{2}-\frac{\omega_+L}{\Delta\xi_0}\right)}\Theta(|\omega_+|-\Delta)\\&
	+\frac{\omega_-^2-\Delta^2}{\omega_-^2-\Delta^2\cos^2\left(\frac{\phi_l}{2}+\frac{\omega_-L}{\Delta\xi_0}\right)}\Theta(|\omega_-|-\Delta).
\end{eqnarray}
The phase difference along the two edges is given by
\begin{eqnarray}
	\phi_u&=\phi_0+\pi\frac{\Phi}{\Phi_0},\\
	\phi_l&=\phi_0-\pi\frac{\Phi}{\Phi_0}.
\end{eqnarray}
The corresponding transmission probabilities for left-moving quasiparticles $\mathcal T_\text{u,L}$, $\mathcal T_\text{l,L}$ are obtained by replacing $\omega_\pm\to\omega_\mp$ and $\phi_\text{u,l}\to-\phi_\text{u,l}$.
Hence, we have $\mathcal T_\text{u,R}+\mathcal T_\text{l,R}=\mathcal T_\text{u,L}+\mathcal T_\text{l,L}$ as expected by the conservation of the charge current carried by quasiparticles in the central part of the junction. Furthermore, the transmission probabilities satisfy $\mathcal T_\text{u,R}(\phi_0,\Phi,p_S)=\mathcal T_\text{u,R}(-\phi_0,-\Phi,-p_S)$ and similarly for the other transmissions in agreement with Onsager symmetry.

Knowing the transmission function, the thermal conductance of the junction can be evaluated straightforwardly as
\begin{equation}
	\kappa(\Phi)=\frac{1}{h}\int \rmd\omega \frac{\omega^2 \mathcal T(\omega)}{4\kBT\cosh^2\frac{\omega}{2\kBT}},
\end{equation}
where we defined $\mathcal T(\omega)=(\mathcal T_\text{u,R}+\mathcal T_\text{l,R})/2$. The factor of two compensates for the double counting that occurs when considering electron- and hole-like quasiparticles at both positive and negative energies. We emphasize that the transmission function $\mathcal T(\omega)$ takes values between zero and two as transport occurs via two counterpropagating edge channels along each edge.

\section{\label{sec:results}Results}
\subsection{Transmission function}
\begin{figure}
	\centering\includegraphics[width=.75\columnwidth]{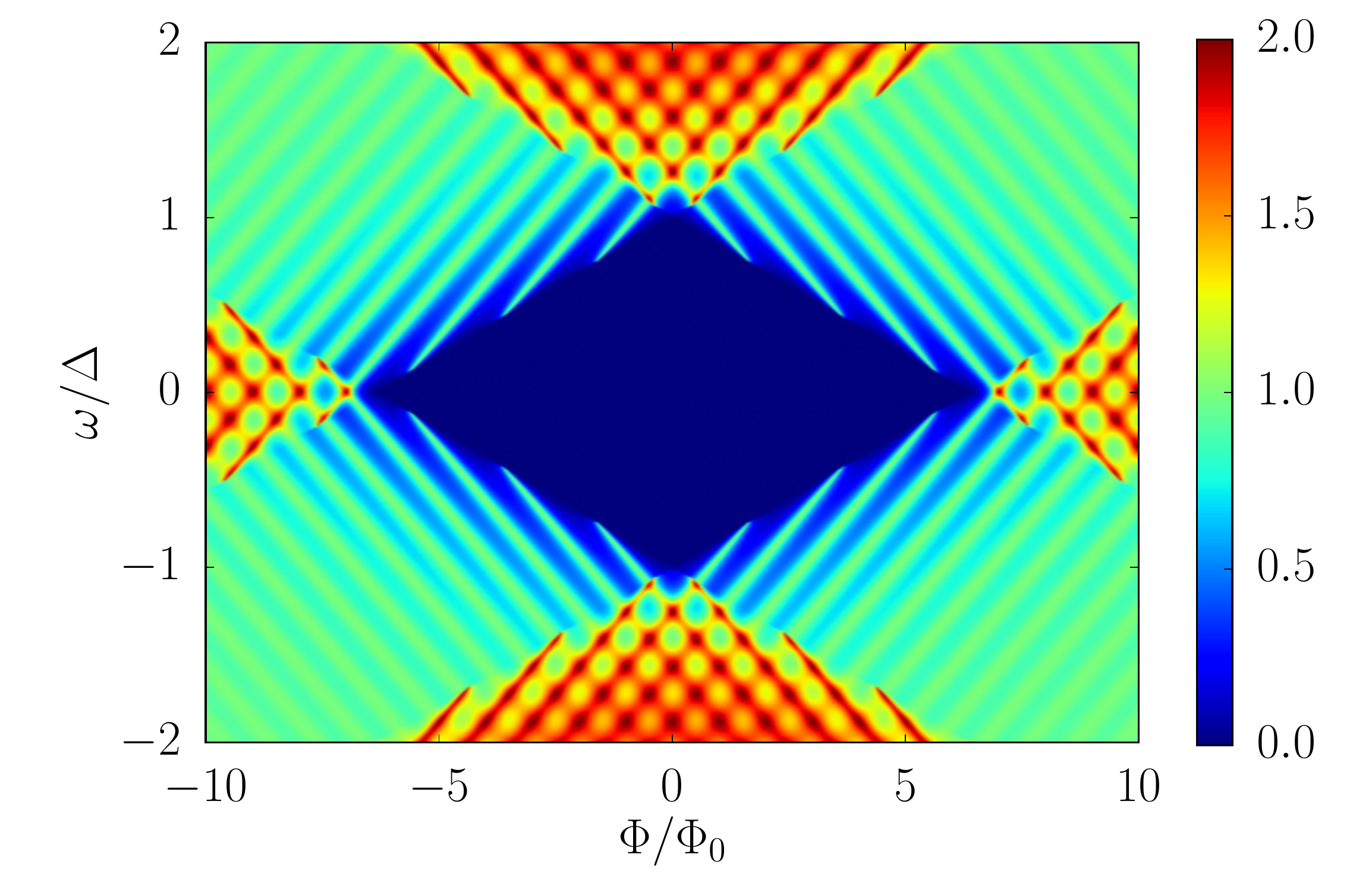}
	\caption{\label{fig:Transmission}Transmission function $\mathcal T(\omega)$ of quasiparticles through a quantum spin Hall Josephson junction. Parameters are $L=10\xi_0$ and $\phi_0=0$. In the blue area (region I), transport is completely blocked due to the superconducting gap. In the green areas (region II), transport occurs via a single channel. In the red areas (region III and IV), transport is possible via two channels.}
\end{figure}

In Fig.~\ref{fig:Transmission}, we show the flux- and energy-dependence of the transmission function $\mathcal T(\omega)$. In total, there are four regions in parameter space with a qualitatively different behavior of the transmission function. 

In region I which occurs for fluxes smaller than $\Phi_*=2L\Phi_0/(\pi\xi_0)$ at small energies, the system is fully gapped. In consequence, quasiparticle transport through the junction is completely blocked and the transmission function vanishes identically, $\mathcal T(\omega)=0$.

Region II occurs for intermediate energies at any flux value. Here, due to the Doppler shift, transport is possible for, say, left-moving quasiparticles while right-moving quasiparticles are still gapped out. As a result, the transmission function takes a maximal value of unity. In addition, in this region $\mathcal T(\omega)$ exhibits an oscillatory behavior. These oscillations can be traced back to the energy dependence of the wave vectors of electrons and holes. Interestingly, the slope of the lines with maximal transmission is twice as large as the slope of the boundaries of the various regions. This is due to an interplay between the flux dependence of $p_S$ and $\phi_{u,l}$ that both enter the oscillatory terms of the transmission function.

In region III, occurring at very large energies for any flux, the transmission function reaches a maximal value of two because transport is enabled now for both left- and right-moving quasiparticles. Finally, in region IV which occurs at small energies for fluxes larger than $\Phi_*$, the Doppler shift is so large that the gap is closed for both left and right movers. As quasiparticle transport is possible along two channels now, the transmission function again reaches a maximal value of two. Just as in region II, both region III and IV exhibit oscillations of $\mathcal T(\omega)$ due to the energy dependence of electron and hole wave vectors.

\subsection{Thermal conductance}
\begin{figure}
	\centering\includegraphics[width=.49\textwidth]{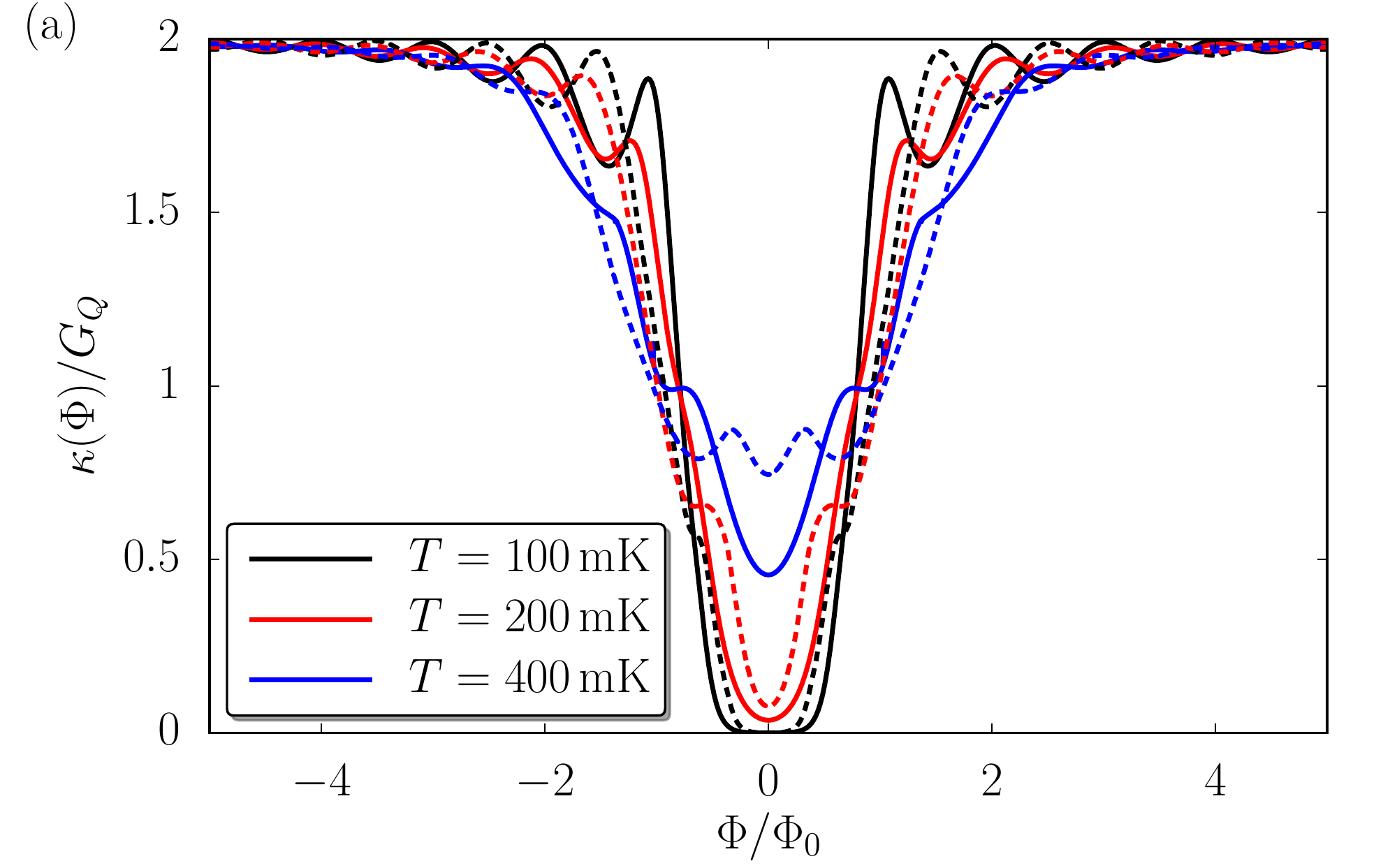}
	\includegraphics[width=.49\textwidth]{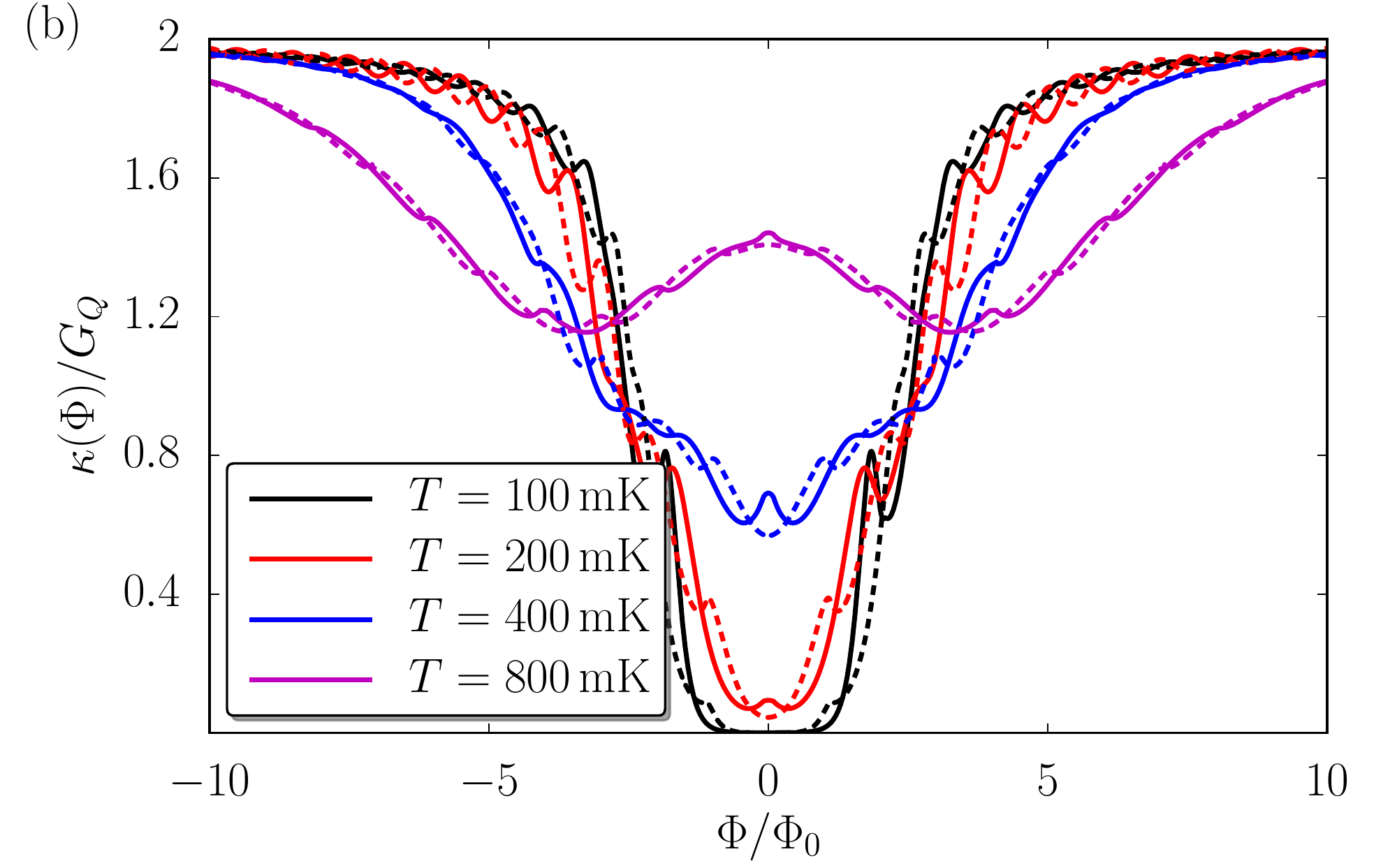}
	\includegraphics[width=.49\textwidth]{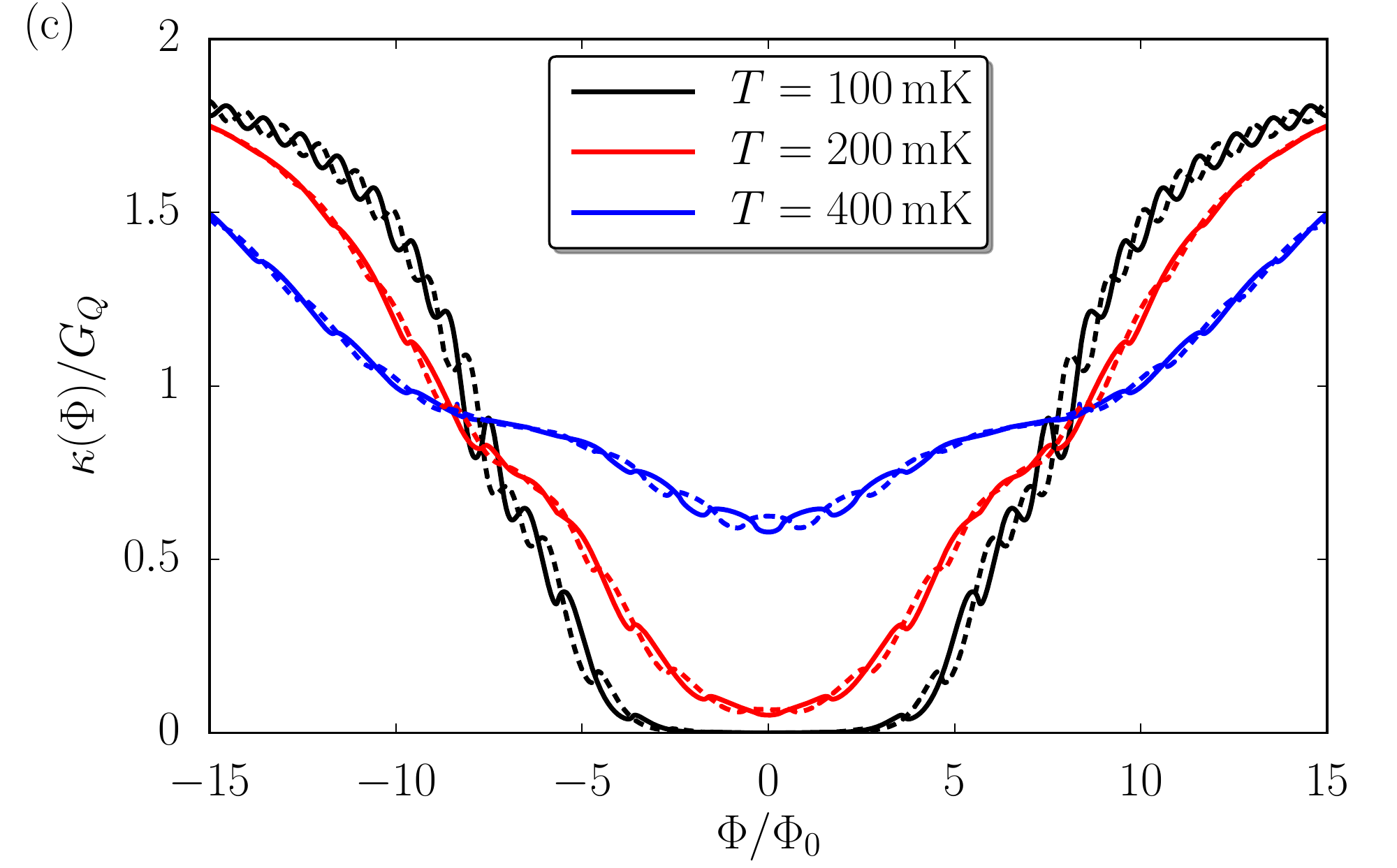}
	\caption{\label{fig:Fraunhofer}Flux-dependence of the thermal conductance for junctions of length (a) $L=\xi_0$, (b) $L=3\xi_0$ and (c) $L=10\xi_0$  at different average temperatures for $\Delta=\unit[0.125]{meV}$. Solid lines correspond to $\phi_0=0$, while dashed lines correspond to $\phi_0=\pi$.}
\end{figure}

We now turn to the thermal conductance of an electrically isolated junction, i.e., a junction where no supercurrent is flowing. This implies that the phase difference in the middle of the junction can take the values $\phi_0=0$ and $\phi_0=\pi$ only. At vanishing magnetic flux, $\Phi=0$, we have $\phi_0=0$. Upon increasing the flux, the system will exhibit phase slips by $\pi$ whenever the Josephson current changes sign~\cite{giazotto_coherent_2013}. In the following, we will discuss the characteristics of the thermal conductance for junctions of three different lengths.

For a short junction, $L=\xi_0$, the thermal conductance is exponentially suppressed at low temperatures and vanishing magnetic flux, cf. Fig.~\ref{fig:Fraunhofer}(a). This is due to the presence of a superconducting gap and a small number of thermally excited quasiparticles, such that transport is dominated by the vanishing transmission function in region I. Upon applying a small magnetic flux $\Phi_*=2/\pi\Phi_0$, the superconducting gap is closed, leading to a large thermal conductance of $\kappa\approx 2G_Q$ where $G_Q=\pi^2\kB^2T/(3h)$ denotes the thermal conductance quantum. The change in thermal conductance is very abrupt and is characterized by a transfer function $\rmd\kappa/\rmd\Phi$ reaching values of up to $4G_Q/\Phi_0$. In addition to the switching behavior, the thermal conductance also shows smaller oscillations which can be traced back to the energy dependence of electron and hole wave vectors.
At higher temperatures, the thermal conductance exhibits a qualitatively similar behavior. However, the suppression of the conductance at $\Phi=0$ is less pronounced due to an increased number of thermally excited quasiparticles.

For junctions of intermediate length, $L=3\xi_0$, an interesting new behavior occurs at elevated temperatures. For $T=\unit[400]{mK}$, the thermal conductance shows a shoulder at $\Phi=\Phi_*$ and $\kappa\approx G_Q$, see Fig.~\ref{fig:Fraunhofer}(b). It occurs because at this point transport is dominated by the finite transmission along a single mode that occurs in region II due to the finite Cooper pair momentum. 
At even higher temperatures, the system can also exhibit a nonmonotonic dependence of the thermal conductance on magnetic flux. This happens since at vanishing flux, the thermal conductance is affected by a vanishing transmission for $\omega<\Delta$ and two open transport channels for thermally excited quasiparticles with $\omega>\Delta$. In contrast, for finite values of $\Phi$, transport becomes determined by a vanishing transmission at small energies and a single open channel at larger energies, thus leading to a reduced thermal conductance.

Finally, for long junctions, $L=10\xi_0$, there is a broad region of exponentially suppressed conductance due to the large value of $\Phi_*$ as shown in Fig.~\ref{fig:Fraunhofer}(c). At the same time, the increase of thermal conductance with $\Phi$ is rather slow. In addition, there is again a shoulder occurring at elevated temperatures for $\Phi=\Phi_*$ and $\kappa\approx G_Q$ due to the same reasons as in the case of a junction of intermediate length.


We remark that the magnetic field required to generate a magnetic flux through the junction breaks time-reversal symmetry. For a junction with an area of $\unit[1]{\mu\textrm{m}^2}$, a flux of $\Phi=\Phi_0$ is generated by a small magnetic field of $\unit[2]{mT}$.
Thus, the topological protection of edge channels gets lifted and backscattering due to disorder and electron-electron interactions can occur. Current experiments have typical superconducting coherence lengths of $\xi_0\approx \unit[600]{nm}$ and mean free paths of a few microns~\cite{bocquillon_gapless_2017,nowack_imaging_2013}. Thus, in order to observe ballistic transport in long junctions, one should reduce the superconducting coherence length by increasing the induced superconducting gap. This is possible by improving the superconductor-topological insulator interfaces or by using superconducting electrodes with larger gaps. Furthermore, an increased mean free path in cleaner samples will help to observe experimentally the effects discussed here for long junctions.

\subsection{Possible experimental setup}
\begin{figure}
	\centering\includegraphics[width=.75\columnwidth]{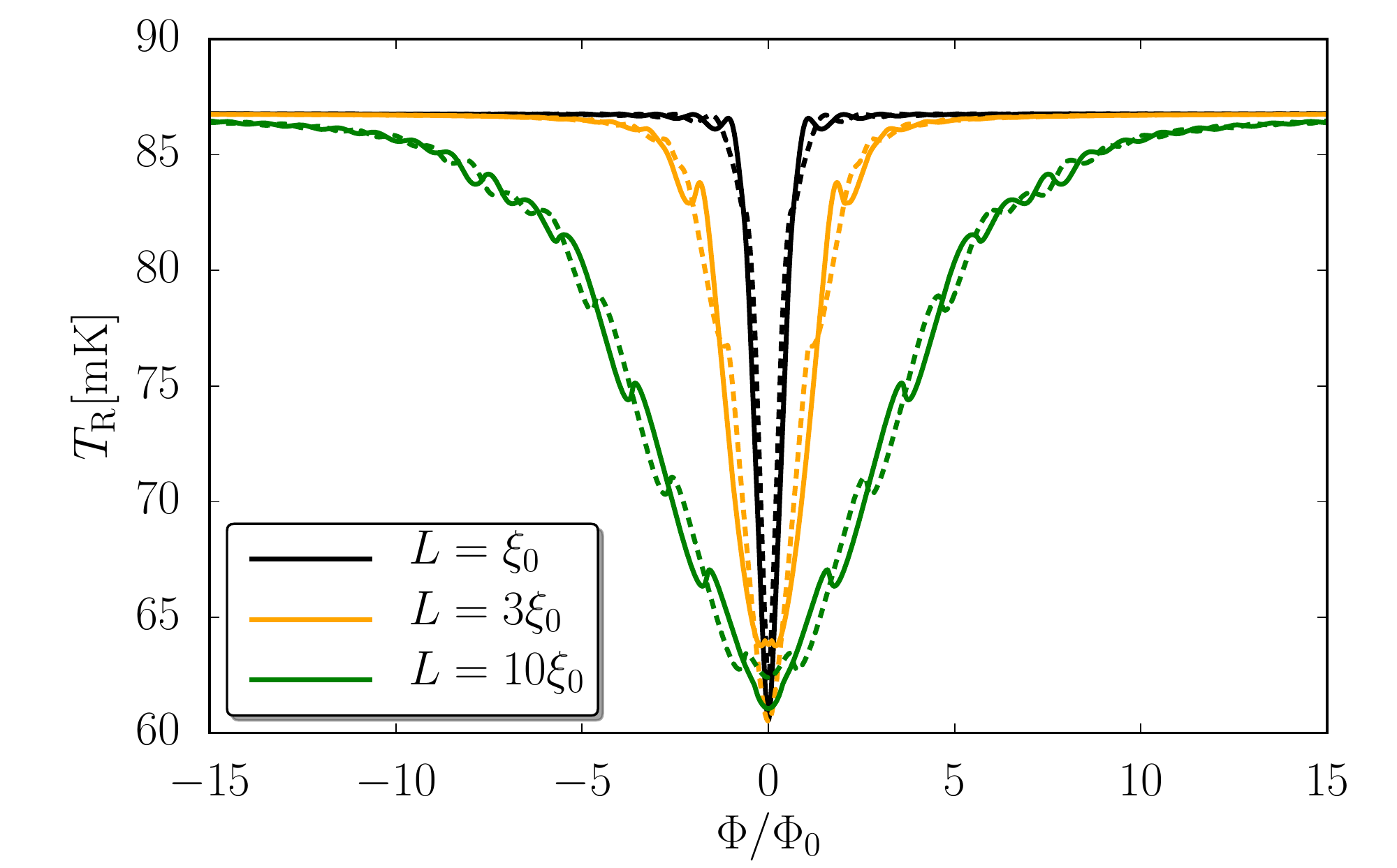}
	\caption{\label{fig:Temperature}Temperature of the right superconducting lead as a function of magnetic flux for different junction lengths. Parameters are $\Sigma\mathcal V=\unit[10^{-3}]{W/K^5}$, $T_\text{bath}=\unit[50]{mK}$, $T_\text{L}=\unit[100]{mK}$, $\Delta=\unit[0.125]{meV}$ and $\phi_0=0$ (solid lines) and $\phi_0=\pi$ (dashed lines), see main text for details.}
\end{figure}

In the following, we discuss a potential experimental setup to realize a phase-coherent thermal switch based on heat diffraction in topological Josephson junctions. 
To this end, we consider a setup as depicted in Fig.~\ref{fig:setup_2D}. The left superconductor can be heated by additional normal metal-insulator-superconductor (NIS)  tunnel junctions while the temperature of the right superconductor can
be measured using similar junctions operated as NIS thermometers~\cite{giazotto_opportunities_2006}.

We account for the heat flows through the system in a simple model.
The heater injects heat at a rate $\dot Q_\text{heat}$. The phase-dependent heat flow between the superconductors through the junction is given by $\dot Q(\Phi)=\kappa(\Phi)(T_\text{L}-T_\text{R})$. 
Finally, heat losses from superconductor $r$ into the substrate lattice due to electron-phonon interactions are modeled phenomenologically as $\dot Q_{\text{e-ph},r}=0.98 \rme^{-\Delta/k_B T_r}\Sigma\mathcal V(T_R^5-T_\text{bath}^5)$ which holds for $T_{\text{bath}}< T_R\ll \Delta/k_B$~\cite{timofeev_recombination-limited_2009}.
Here, $\Sigma$ denotes the material-dependent electron-phonon coupling~\cite{giazotto_opportunities_2006}, $\mathcal V$ is the volume of the superconducting electrode, and $T_\text{bath}$ is the temperature of the phonon bath. In the stationary state, the sum of the heat currents flowing in and out of each superconducting electrode has to vanish. This allows us to determine the flux-dependent temperature of the right superconductor for a fixed temperature of the left electrode.

In order to achieve a large switching effect, for $\Phi=0$ the temperature of the right superconductor should be as low as possible. This requires the electron-phonon coupling to dominate over the electronic heat current through the junction. As both these quantities are exponentially suppressed by the superconducting gap (which is of the order of $\Delta\sim\unit[0.1]{meV}$ in recent experiments on HgTe~\cite{bocquillon_gapless_2017,wiedenmann_4-periodic_2016} and InAs/GaSb~\cite{pribiag_edge-mode_2015}) at low temperatures, this necessitates that $\Sigma\mathcal V$ is sufficiently large.
At the same time, for a large magnetic flux $\Phi$, the temperature of the right superconductor should be as large as possible to realize a good switch. The right superconductor should thus thermalize with the heated left superconductor. Hence, the temperatures of the superconductors should be much smaller than the superconducting gap such that the electron-phonon coupling is exponentially small and, thus, overcome by  the electrical thermal conductance which is finite due to the magnetic closing of the gap. Indeed, as seen in Fig.~4, $\kappa(\Phi)$ increases with the magnetic flux which allows for tuning of the ratio of electron-phonon interactions in respect to  the electrical thermal conductance, and therefore a large thermal switching effect between $\Phi=0$ and large $\Phi$ limits.

The resulting behavior of the thermal switch is displayed in Fig.~\ref{fig:Temperature}. It shows that for realistic system parameters, a large temperature switching of about $\unit[25]{mK}$ can be achieved. This value turns out to be  much larger than the current achievable experimental temperature resolution of $\unit[0.1]{mK}$~\cite{giazotto_opportunities_2006}, and corresponds to a relative temperature variation $\mathcal{R}=(T_R^\text{max}-T_R^\text{min})/T_R^\text{min}\sim 40\%$, where $T_R^\text{max}$($T_R^\text{min}$) is the maximum(minimum) temperature obtained with the switch. 
The above ratio can be compared to the ones obtained so far in  Josephson coherent thermal modulators realized with conventional superconducting tunnel junctions where $\mathcal{R}$ values up to $\sim 20\%$ have been measured~\cite{fornieri_nanoscale_2016}, 
and reveals how Josephson junctions based on two-dimensional topological insulators can be effective for the implementation of high-performance phase-tunable thermal switches.

We also emphasize that the effect is robust with respect to a variation of parameters.
Notably, the magnitude of the temperature change is independent of the junction length. However, the junction length has a crucial impact on the switching characteristics. In the short junction limit, there is a very abrupt switching behavior, while for long junction, the switching occurs smoothly over a larger flux range. The latter behavior can be advantageous to finely tune the temperature of a superconducting electrode whereas the former behavior provides a very sensitive switch.

So far, we have analyzed an ideal situation where transport occurs exclusively via edge channels. 
Whilst Josephson junctions where transport was indeed exclusively due to edge channel transport have been successfully fabricated, there is also the possibility that in a given sample transport occurs both along the edge as well as via the lowest bulk mode. In the following, we will argue that the presence of bulk transport will have no significant impact on the switch performance.
At  vanishing magnetic flux, both the edge channels as well as the bulk mode are gapped due to the presence of superconducting electrodes. Hence, transport is exponentially suppressed even in the presence of bulk modes. Upon increasing the magnetic flux, the finite Cooper pair momentum $p_S$ will close the gap for the edge channels once $\Phi>\Phi_*$, thus leading to a thermal conductance $\kappa\approx 2G_Q$. Importantly, the bulk mode which is localized at the center of the junction feels a much smaller Cooper pair momentum and is thus still gapped at $\Phi=\Phi_*$, i.e., it only has an exponentially suppressed contribution to the thermal conductance, and thereby does not affect the thermal switching behavior.

\section{\label{sec:conclusions}Conclusions}
We demonstrated that the diffraction of coherent heat currents in topological Josephson junctions can be used to realize an effective thermal switch with relative temperature variations up to $\sim 40\%$ between the on and off state.
The switch is controlled by the magnetic flux through the junction which provides Cooper pairs with a finite momentum. The latter allows for a controlled opening and closing of the superconducting gap.
For short junctions, we find a sharp switching behavior whereas for long junctions there is a continuous switching. Junctions of intermediate length can exhibit a nonmonotonic flux dependence of the thermal conductance.
Our proposal is within the reach of current experimental technology, and can provide an important building block for future phase-coherent caloritronic devices which may be potentially used for thermal logic and management at the nanoscale.

\ack
B.S. and E.M.H. acknowledge financial support from the DFG via SFB 1170 "ToCoTronics", and the ENB Graduate School on Topological Insulators.
B.S. acknowledges financial support from the Ministry of Innovation NRW.
F.G. acknowledges the MIUR-FIRB2013-Project Coca (grant no. RBFR1379UX), and the European Research Council under the European
Union's Seventh Framework Programme (FP7/2007-2013)/ERC grant agreement no. 615187 - COMANCHE for partial financial support.

\section*{References}


\providecommand{\newblock}{}

\end{document}